\documentclass[12pt]{article}
\def\int {\intop \limits}

\def\fnote#1{\footnote}
\begin{document}

\renewcommand \theequation{\thesection.\arabic{equation}}

\title{Radiation from polarized electrons in oriented crystals 
at high energy}
\author{V. N. Baier
and V. M. Katkov\\
Budker Institute of Nuclear Physics\\ 630090 Novosibirsk,
Russia}

\maketitle

\begin{abstract}
Radiation from high energy electrons in an oriented crystal can be 
considered in a frame of the quasiclassical operator method 
which appears to be a most satisfactory approach to the problem. 
Under some quite generic assumptions the general expression is derived
for the probability of circularly polarized photon emission 
from the longitudinally polarized electron in oriented crystal.
The particular mechanism of radiation depends
on interrelation between the angle of incidence $\vartheta_0$ (angle
between the momentum of initial electron and axis (plane) of crystal) and
angle $\vartheta_v \equiv V_0/m$~($V_0$ is the scale of a potential
of axis or a plane relative to which the angle $\vartheta_0$ is defined). 
When $\vartheta_0 \ll \vartheta_v$
one has magnetic bremsstrahlung type of radiation (with corrections 
$\propto \vartheta_0^2$ which are due to inhomogeneous character of field in
crystal). When $\vartheta_0 \gg \vartheta_v$ one obtains the theory of
coherent bremsstrahlung, while for $\vartheta_0 \geq \vartheta_v$ one 
arrives to the modified theory of coherent bremsstrahlung. At high energy
radiation in oriented crystals is strongly enhanced comparing with standard 
bremsstrahlung.

\end{abstract}

\newpage

\section{Introduction}

The quasiclassical operator method developed by 
authors \cite{BK0}-\cite{BK2} is adequate for consideration of the 
electromagnetic processes at high energy.
The probability of photon emission has a form (see
\cite{BKS}, p.63, Eq.(2.27); the method is given also in
\cite{BLP},\cite{BKF})
\begin{equation}
dw=\frac{e^2}{(2\pi)^2} \frac{d^3k}{\omega}
\int_{}^{}dt_2\int_{}^{}dt_1 R^{\ast}(t_2) R(t_1)
\exp \left[-\frac{i\varepsilon}{\varepsilon'}
\left(kx(t_2)-kx(t_1)\right) \right],
\label{1}
\end{equation}
where $k^{\mu}=(\omega, {\bf k})$ is the 4-momentum of the emitted
photon, $k^2=0$, $x^{\mu}(t)=(t, {\bf r}(t))$, $t$ is the time,
and ${\bf r}(t)$ is the particle location on a classical trajectory,
$kx(t)=\omega t- {\bf kr}(t)$,
$\varepsilon'=\varepsilon-\omega$, 
we employ units such that $\hbar=c=1$.
The matrix element $R(t)$ is defined by the structure of a current.
For an electron (spin 1/2 particle) one has
\begin{eqnarray}
&& R(t)=\frac{m}{\sqrt{\varepsilon \varepsilon'}}
\overline{u}_{s_{f}}({\bf p'})\hat{e}^{\ast}u_{s_{i}}({\bf p})
=\varphi_{s_{f}}^{+}\left(A(t)+i\mbox{\boldmath$\sigma$}{\bf B}(t) \right)
\varphi_{s_{i}}, 
\nonumber \\
&& A(t)= \frac{1}{2}\left(1+\frac{\varepsilon}{\varepsilon'} \right)
{\bf e}^{\ast}\mbox{\boldmath$\vartheta$}(t), 
\nonumber \\
&& {\bf B(t)}=\frac{\omega}{2\varepsilon'}\left({\bf e}^{\ast} \times
\left(\frac{{\bf n}}{\gamma}- \mbox{\boldmath$\vartheta$}(t)\right) \right),
\label{2}
\end{eqnarray}
here  ${\bf e}$ is the vector of the polarization of a photon
(the Coulomb gauge is used),the four-component spinors $u_{s_f},
u_{s_i}$ and the two-component spinors $\varphi_{s_f},
\varphi_{s_i}$ describe the initial ($s_i$) and final ($s_f$)
polarization of the electron, {\bf v}={\bf v}(t) is the electron
velocity, $\mbox{\boldmath$\vartheta$}(t) = \left({\bf v}-{\bf
n}\right) \simeq {\bf v}_{\perp}(t)$, ${\bf
v}_{\perp}$ is the component of particle velocity perpendicular
to the vector ${\bf n}={\bf k}/|{\bf k}|$, $\gamma=\varepsilon/m$
is the Lorentz factor. The expressions in Eq.(\ref{2}) are
given for radiation of ultrarelativistic electrons, they are
written down with relativistic accuracy (terms $\sim 1/\gamma$
are neglected) and in the small angle approximation.

The important parameter $\chi$
characterizes the quantum effects in an external field, when
$\chi \ll 1$ we are in the classical domain and with $\chi \geq
1$ we are already well inside the quantum domain
\begin{equation}
\chi =\frac{|{\bf F}|\gamma}{F_0}, \quad {\bf F}={\bf
E}_{\perp}+({\bf v}\times {\bf H}),\quad {\bf E}_{\perp}={\bf E}-
{\bf v}({\bf v}{\bf E}),
\label{3}
\end{equation}
where ${\bf E} ({\bf H})$ is an electric (magnetic) field,
$F_0=m^2/e=(m^2c^2/e\hbar)$ is the quantum boundary (Schwinger)
field: $H_0=4.41 \cdot 10^{13} {\rm Oe},~ E_0=1.32\cdot 10^{16} 
{\rm V/cm}$. 

The quasiclassical operator method is applicable when $H \ll H_0,~
E \ll E_0$ and $\gamma \gg 1$.

The general expression for combination $R^{\ast}(t_2) R(t_1)=R_2^{\ast}R_1$
with all polarization taken into account is
\begin{equation}
R_2^{\ast}R_1 \equiv N_{21}(\mbox{\boldmath$\zeta$}_i, 
\mbox{\boldmath$\zeta$}_f, \textbf{e})=\frac{1}{4} 
{\rm Tr}\left[(1+\mbox{\boldmath$\zeta$}_i \mbox{\boldmath$\sigma$})
\left(A_2^{\ast}-i \mbox{\boldmath$\sigma$}\textbf{B}_2^{\ast} \right)
(1+\mbox{\boldmath$\zeta$}_f \mbox{\boldmath$\sigma$})  
\left(A_1+i \mbox{\boldmath$\sigma$}\textbf{B}_1 \right)\right],
\label{4} 
\end{equation} 
where we use for the description of electron polarization 
the vector  $\mbox{\boldmath$\zeta$}$  describing the
polarization of the electron (in its rest frame),
$\mbox{\boldmath$\zeta$}_i$ is the spin vector of initial electron,
$\mbox{\boldmath$\zeta$}_f$ is the spin vector of final electron.
Summing over final spin states we have
\begin{equation}
\sum_{\zeta_f}R_2^{\ast}R_1 =A_2^{\ast}A_1
+\textbf{B}_2^{\ast}\textbf{B}_1
+i\left[A_2^{\ast}(\mbox{\boldmath$\zeta$}_i\textbf{B}_1)
-A_1  (\mbox{\boldmath$\zeta$}_i\textbf{B}_2^{\ast})
+\mbox{\boldmath$\zeta$}_i\left(\textbf{B}_2^{\ast}\times
\textbf{B}_1\right) \right],  
\label{5} 
\end{equation} 
where the two first terms describe the radiation of unpolarized 
electrons and the last terms is an addition dependent on 
the initial spin.

A combination of matrix elements $R_2^{\ast}R_1$ for spin-flip 
transition follows from Eq.(\ref{4}) after substitution 
$\mbox{\boldmath$\zeta$}_i =\mbox{\boldmath$\zeta$}$,
$\mbox{\boldmath$\zeta$}_f=-\mbox{\boldmath$\zeta$}$
\begin{equation}
N_{21}(\mbox{\boldmath$\zeta$}, -\mbox{\boldmath$\zeta$}, \textbf{e})=
\textbf{B}_2^{\ast}\textbf{B}_1-(\mbox{\boldmath$\zeta$}\textbf{B}_2^{\ast})
(\mbox{\boldmath$\zeta$}\textbf{B}_1)-i\mbox{\boldmath$\zeta$}
\left(\textbf{B}_2^{\ast}\times\textbf{B}_1\right).
\label{6} 
\end{equation} 
It depends on the spin-flip amplitudes $\textbf{B}_{1,2}$ only. 

After summing in Eq.(\ref{5}) over the polarization of emitted photon
$\lambda$  we obtain for unpolarized electrons
\begin{equation}
\sum_{\lambda, \zeta_f} R_2^{\ast} R_1 \rightarrow \frac{1}{2\varepsilon'^2}
\left[\frac{\omega^2}{\gamma^2}+\left(\varepsilon^2+\varepsilon'^2 \right)
\mbox{\boldmath$\vartheta$}_1\mbox{\boldmath$\vartheta$}_2 \right].
\label{7}
\end{equation}
For the longitudinally polarized initial electron and for circular 
polarization of emitted photon we have
\begin{equation}
\sum_{\lambda, \zeta_f} R_2^{\ast}R_1 = \frac{1}{2\varepsilon'^2}
\left\{\frac{\omega^2}{\gamma^2}(1+\xi)+
\left[(1+\xi)\varepsilon^2+(1-\xi)\varepsilon'^2
\right]\mbox{\boldmath$\vartheta$}_1\mbox{\boldmath$\vartheta$}_2 \right\},
\label{8}
\end{equation}
where $\xi=\lambda\zeta,~\lambda=\pm 1$ is the helicity of emitted photon,
$\zeta =\pm 1$ is the helicity of the initial electron. 
In this expression we omit the terms which vanish after integration
over angles of emitted photon.

Just for selected polarizations there are the strong effects 
in the radiation probability in high-energy limit: in the hard part 
of spectrum the longitudinally polarized initial electron emits 
mainly circularly polarized photon. This is a particular example of
helicity transfer.

\section{General approach to radiation in oriented crystals}
\setcounter{equation}{0}

The theory of high-energy electron radiation and electron-positron pair
creation in oriented crystals was developed in \cite{BKS1}-\cite{BKS2},
and given in \cite{BKS}. In these publications the radiation
from unpolarized electrons was considered including 
the polarization density matrix
of emitted photons. Since Eq.(\ref{8}) has the same structure as
Eq.(\ref{7}), below we use systematically the methods of mentioned papers
to obtain the characteristics of radiation 
from longitudinally polarized electron.

Let us remind that along with the parameter $\chi$ which 
characterizes the quantum
properties of radiation there is another parameter
\begin{equation}
\varrho=2\gamma^2\left\langle (\Delta\textbf{v} )^2\right\rangle,
\label{1.1}
\end{equation} 
where $\left\langle (\Delta\textbf{v} )^2\right\rangle=
\left\langle \textbf{v}^2 \right\rangle -\left\langle \textbf{v} \right\rangle^2$
and $\left\langle\ldots \right\rangle$ denotes averaging over time.
In the case $\varrho \ll 1$ the radiation is of a dipole nature and it is formed
during the time of the order of the period of motion. In the case 
$\varrho \gg 1$ the radiation is of magnetic bremsstrahlung nature and
it is emitted from a small part of the trajectory.

In a crystal the parameter $\varrho$ depends on the angle incidence $\vartheta_0$
which is the angle between an axis (a plane) of crystal and the momentum
of a particle. If $\vartheta_0 \leq \vartheta_c$ (where $\vartheta_c \equiv
(2V_0/\varepsilon)^{1/2},~V_0$ is the scale of continuous potential of 
an axis or a plane relative to which the angle $\vartheta_0$ is defined)
electrons falling on a crystal are captured into channels or low above-barrier
states, whereas for $\vartheta_0 \gg \vartheta_c$ the incident particles 
move high above the barrier. In later case we can describe the motion using 
the approximation of the  rectilinear trajectory, 
for which we find from Eq.(\ref{1.1}) the
following estimate
\begin{equation}
\varrho(\vartheta_0)=\left(2V_0/m\vartheta_0 \right)^2,
\label{2.1}
\end{equation} 
For angles of incidence in the range $\vartheta_0 \leq \vartheta_c$
the transverse (relative to an axis or a plane) velocity of particle
is $v_{\perp}\sim \vartheta_c$ and the parameter obeys $\varrho \sim \varrho_c$
where
\begin{equation}
\varrho_c=2V_0\varepsilon/m^2.
\label{3.1}
\end{equation} 
This means that side by side with the Lindhard angle $\vartheta_c$ the
problem under consideration has another characteristic angle
$\vartheta_v=V_0/m$ and $\varrho_c=(2\vartheta_v/\vartheta_c)^2$.

We consider here the photon emission in a thin crystal  when the condition 
$\varrho_c \gg 1$ is satisfied.
In this case the extremely difficult task of averaging of Eq.(\ref{1}) and
Eqs.(\ref{4})-(\ref{8}),  derived for a given trajectory,  over all 
possible trajectories of electrons in a crystal  simplifies radically.
In fact, if $\varrho_c \gg 1$ then in the range where trajectories are
essentially non-rectilinear ($\vartheta_0 \leq \vartheta_c,~v_{\perp} \sim 
\vartheta_c$) the mechanism of photon emission is of the magnetic 
bremsstrahlung nature and the characteristics of radiation can be expressed
in terms of local parameters of motion. Then the averaging procedure can be  
carried out simply if one knows the distribution function in the transverse
phase space $dN(\varrho, \textbf{v}_{\perp})$, which for a thin crystal 
is defined directly by the initial conditions of incidence of particle
on a crystal. For a given angle of incidence $\vartheta_0$ we have
$dN/N=d^3r F(\textbf{r}, \vartheta_0)/V$, where $V$ is the volume of a 
crystal and $N$ is the total number of particles. In the axial case
the function $F(\textbf{r}, \vartheta_0)$ is of the form
\begin{equation}
F_{ax}(\mbox{\boldmath$\varrho$}, \vartheta_0)=\int 
\frac{d^2 \varrho_0}{S(\varepsilon_{\perp}(\mbox{\boldmath$\varrho$}_0))}
\vartheta((\varepsilon_{\perp}(\mbox{\boldmath$\varrho$}_0)-
U(\mbox{\boldmath$\varrho$})),
\label{4.1}
\end{equation} 
where $U(\mbox{\boldmath$\varrho$})$ is the continuous axial potential
dependent on the transverse coordinate $\mbox{\boldmath$\varrho$}$ normalized 
so that $U(\mbox{\boldmath$\varrho$})=0$ at the boundary of a cell;
\begin{equation}
S(\varepsilon_{\perp}(\mbox{\boldmath$\varrho$}_0))=
\int \vartheta((\varepsilon_{\perp}(\mbox{\boldmath$\varrho$}_0)-
U(\mbox{\boldmath$\varrho$}))d^2\varrho,\quad
\varepsilon_{\perp}(\mbox{\boldmath$\varrho$}_0)=
\frac{\varepsilon \vartheta_0^2}{2} +
U(\mbox{\boldmath$\varrho$}_0),
\label{5.1}
\end{equation} 
where $\vartheta(x)$ is the Heaviside function:$\vartheta(x)=0$ if $x < 0$
and $\vartheta(x)=1$ if $x > 0$.
For the planar case we have
\begin{equation}
F_{pl}(x, \vartheta_0)=\int \frac{\vartheta(\varepsilon_{\perp 0}-U(x))dx_0}
{v(\varepsilon_{\perp 0}, x)}
\left[\int \frac{\vartheta(\varepsilon_{\perp 0}-U(y))dy}
{v(\varepsilon_{\perp 0}, y)} \right]^{-1}, 
\label{6.1}
\end{equation} 
where $U(x)$ is the continuous potential of plane dependent on the 
coordinate $x$,~$v(\varepsilon_{\perp 0}, x)=
\left[2(v(\varepsilon_{\perp 0}-U(x))/\varepsilon \right]^{1/2}$
is the transverse velocity of an electron, and $\varepsilon_{\perp 0}
=\varepsilon \vartheta_0^2/2+U(x_0)$. It should be noted that in the
case when $\varepsilon_{\perp}(\mbox{\boldmath$\varrho$}_0)>U_0$
(the above-barrier particles) the distribution Eq.(\ref{4.1}) becomes
uniform, whereas the distribution Eq.(\ref{6.1}) becomes
uniform only if $\varepsilon_{\perp 0} \gg U_0$, $U_0$ is the depth
of potential well.

Substituting Eq.(\ref{8}) into  Eq.(\ref{1}) we find after 
integration by parts of terms ${\bf nv_{1,2}}~({\bf nv_{1,2}} \rightarrow 1)$
the general expression for energy loss of radiating electron 
($dE_{\xi} = \omega dw_{\xi}$) 
\begin{eqnarray}
&& dE_{\xi} = -\frac{\alpha m^2}{8 \pi^2}\frac{d^3k}{\varepsilon\varepsilon'}
\int \frac{d^3r}{V}F(\textbf{r}, \vartheta_0) \int e^{-iA}
\left[\varphi_1(\xi)+\frac{1}{4}\varphi_2(\xi)\gamma^2\left(\textbf{v}_1-
\textbf{v}_2\right)^2\right]dt_1dt_2,
\nonumber \\
&& A=\frac{\omega \varepsilon}{2\varepsilon'}\int_{t_1}^{t_2}
\left[\frac{1}{\gamma^2}+(\textbf{n}-\textbf{v}(t))^2\right]dt,
\nonumber \\
&&\varphi_1(\xi)=1+\xi\frac{\omega}{\varepsilon}, \quad
\varphi_2(\xi)=(1+\xi)\frac{\varepsilon}{\varepsilon'}+
(1-\xi)\frac{\varepsilon'}{\varepsilon}.
\label{7.1}
\end{eqnarray}
where $\alpha=e^2=1/137$, the vector $\textbf{n}$ is defined 
in Eq.(\ref{1}), the helicity of emitted photon $\xi$
is defined in Eq.(\ref{8}).

The circular polarization of radiation is defined by the Stoke's parameter 
$\xi^{(2)}$:
\begin{equation}
\xi^{(2)}=\Lambda (\mbox{\boldmath$\zeta$}{\bf v}),\quad
\Lambda = \frac{dE_+ - dE_-}{dE_+ + dE_-},   
\label{7.1a}
\end{equation} 
where the quantity $(\mbox{\boldmath$\zeta$}{\bf v})$ defines the longitudinal
polarization of the initial electrons, $dE_+$ and $dE_-$ is the energy
loss for $\xi$=+1 and $\xi$=-1 correspondingly. In the limiting case
$\omega \ll \varepsilon$ one has $\varphi_2(\xi) \simeq 2(1+\xi\omega/\varepsilon)
=2\varphi_1(\xi)$. So the expression for the energy loss $dE_{\xi}$ contains the
dependence on $\xi$ only as a common factor $\varphi_1(\xi)$ only. Substituting
in Eq.(\ref{7.1a}) one obtains the universal result independent of a particular
mechanism of radiation
\begin{equation}
\xi^{(2)}=\frac{\omega}{\varepsilon}(\mbox{\boldmath$\zeta$}{\bf v})
\label{7.2a}
\end{equation} 

The periodic crystal potential $U(\textbf{r})$ can be
presented as the Fourier series (see e.g.\cite{BKS}, Sec.8 )
\begin{equation}
U(\textbf{r})=\sum_{\textbf{q}}G(\textbf{q})e^{-i\textbf{q}\textbf{r}},
\label{8.1}
\end{equation} 
where $\textbf{q}=2\pi(n_1, n_2, n_3)/l;~l$ is the lattice constant.
The particle velocity can be presented in a form
$\textbf{v}(t) =\textbf{v}_0+\Delta\textbf{v}(t)$, where 
$\textbf{v}_0$ is the average velocity. If 
$\vartheta_0 \gg \vartheta_c$, we find $\Delta\textbf{v}(t)$
using the rectilinear trajectory approximation 
\begin{equation}
\Delta\textbf{v}(t)=-\frac{1}{\varepsilon}\sum 
\frac{G(\textbf{q})}{q_{\parallel}}\textbf{q}_{\perp}
\exp [-i(q_{\parallel}t+\textbf{q}\textbf{r})],
\label{9.1}
\end{equation} 
where $ q_{\parallel}=(\textbf{q}\textbf{v}_0),~
\textbf{q}_{\perp}=\textbf{q}-\textbf{v}_0(\textbf{q}\textbf{v}_0)$.
If $\varrho_c \gg 1$, there is a range of angles $\vartheta_0$ which
satisfies $\vartheta_c \ll \vartheta_0 \ll \vartheta_v$, where
both the magnetic bremsstrahlung description and the rectilinear 
trajectory approximation are valid. However, since the magnetic 
bremsstrahlung approach provides the same description for all 
angles $\vartheta_0 \ll \vartheta_v$, the formula obtained in some
range of $\vartheta_0$ remain valid up to $\vartheta_0=0$. It is 
known that in range $\vartheta_0 \geq \vartheta_v$ 
the rectilinear trajectory approximation is valid as well for 
$\varrho_c \gg 1$. So, after integration in Eq.(\ref{7.1}) over
${\bf u}={\bf n}-{\bf v}_0~(d^3k=\omega^2 d\omega d{\bf u})$ using 
Eq.(\ref{9.1})
and passing to the variables $t, \tau:~t_1=t-\tau,~t_2=t+\tau$, 
we obtain after simple calculations the general expression for       
the intensity of radiation valid for any angle of incidence 
$\vartheta_0$
\begin{eqnarray}
&& dI_{\xi} \equiv \frac{dE_{\xi}}{dt} 
= \frac{i\alpha m^2}{4 \pi\varepsilon^2}
\omega d\omega \int \frac{d^3r}{V}F(\textbf{r}, \vartheta_0) 
\int \frac{d\tau}{\tau-i0}\Bigg[\varphi_1(\xi)-\varphi_2(\xi)
\nonumber \\
&& \times \left(\sum_{\textbf{q}} 
\frac{G(\textbf{q})}{m q_{\parallel}}\textbf{q}_{\perp}
\sin(q_{\parallel}\tau)e^{-i\textbf{q}\textbf{r}} \right)^2 \Bigg]
  e^{-iA_2},
\label{10.1}
\end{eqnarray}
where 
\begin{eqnarray}
&& A_2=\frac{m^2\omega\tau}{\varepsilon\varepsilon'} \left[1+
\sum_{\textbf{q},\textbf{q}^{\prime} }\frac{G(\textbf{q})G(\textbf{q}')}
{m^2 q_{\parallel} q_{\parallel}'}
(\textbf{q}_{\perp}\textbf{q}_{\perp}')\Psi(q_{\parallel}, q_{\parallel}', \tau)
\exp [-i(\textbf{q}+\textbf{q}')\textbf{r}]\right] 
\nonumber \\
&& \Psi(q_{\parallel}, q_{\parallel}', \tau)=
\frac{\sin(q_{\parallel}+ q_{\parallel}')\tau}{(q_{\parallel}+ 
q_{\parallel}')\tau}-\frac{\sin(q_{\parallel}\tau)}{q_{\parallel}\tau}
\frac{\sin(q_{\parallel}'\tau)}{q_{\parallel}'\tau}.
\label{11.1}
\end{eqnarray}

\section{Radiation for $\vartheta_0 \ll V_0/m$ 
(magnetic \\ bremsstrahlung limit)}
\setcounter{equation}{0}

The behavior of intensity $I$ Eq.(\ref{10.1}) for various entry angles and
energies is determined by the dependence on these parameters of the phase $A_2$
given Eq.(\ref{11.1}). Here we consider the axial case for
$\vartheta_0 \ll V_0/m \equiv \vartheta_v$. The direction of crystal axis we take as
$z-$axis of the coordinate system. The order of magnitude of the double sum in $A_2$
is $(G/m)^2(q_{\perp}/q_{\parallel})^2\Psi(q_{\parallel}, q_{\parallel}', \tau)$.
For the vector $\textbf{q}$ lying in the plane $(x,y)$ we introduce notation
$\textbf{q}_t$, for such vectors one has $q_z=0$ and the quantities in 
Eq.(\ref{11.1}) can be estimated in the following way:
\begin{equation}
G(\textbf{q}) \sim V_0,\quad q_{\perp} \sim 1/a,\quad q_{\parallel} \sim \vartheta_0/a,
\label{1.2}
\end{equation}
where $a$ is the size of the region of action of the continuous potential. For all 
remaining vectors $q_{\perp} \sim q_{\parallel} \sim 1/a$. Then the contribution 
to the sum of the terms with $q_z \neq 0$ will be $\sim (V_0/m)^2\Psi \leq (V_0/m)^2$.
Since $(V_0/m)^2 \ll 1$ this contribution can be neglected. Thus, we keep in the sum
only terms with $\textbf{q}_t$ for which its value is $\sim (V_0/m\vartheta_0)^2\Psi$.
The large value of the phase $A_2$ leads to an exponential suppression of intensity
$I$. Therefore the characteristic value of the variable $\tau$ in the integral
Eq.(\ref{10.1}) (which have the meaning of the formation time (length) of the process)
will be adjusted in a such way that the large factor $(V_0/m\vartheta_0)^2$ will be
compensated by the function $\Psi(q_{\parallel}, q_{\parallel}', \tau)$, i.e. for
small entry angles the contribution gives region where $q_{\parallel}\tau \ll 1$.
Expanding the phase $A_2$ correspondingly we find an approximate expression for
$\vartheta_0 \ll \vartheta_v$ 
\begin{eqnarray}
&& A_2 \simeq \frac{m^2\omega\tau}{\varepsilon\varepsilon'}\Bigg\{1-
\frac{\tau^2}{3} \sum_{\textbf{q}_t,\textbf{q}_t'}
G(\textbf{q}_t)G(\textbf{q}_t')\frac{(\textbf{q}_t\textbf{q}_t')}{m^2}
\exp\left[-i(\textbf{q}_t+\textbf{q}_t')\mbox{\boldmath$\varrho$} \right]
\nonumber \\
&& \times \left[ 1-\frac{\tau^2}{10}\left((\textbf{q}_t\mbox{\boldmath$\nu$})^2
+ (\textbf{q}_t'\mbox{\boldmath$\nu$})^2
+\frac{2}{3} (\textbf{q}_t\mbox{\boldmath$\nu$})
(\textbf{q}_t'\mbox{\boldmath$\nu$})\right) \right]\Bigg\},
\label{2.2}  
\end{eqnarray} 
here $\mbox{\boldmath$\nu$}=\textbf{v}_0/v_0$ is the direction of entry of the 
initial electron, $\mbox{\boldmath$\varrho$}=\textbf{r}_t$. We cam rewrite 
Eq.(\ref{2.2}) in the terms of the average potential of atomic string
$\displaystyle{U(\mbox{\boldmath$\varrho$})=\sum_{\textbf{q}_t} G(\textbf{q}_t)
\exp(-i\textbf{q}_t \mbox{\boldmath$\varrho$})}$:
\begin{equation}
A_2 = \frac{m^2\omega\tau}{\varepsilon\varepsilon'}
\left\lbrace 1+\frac{\tau^2}{3} \textbf{b}^2\tau^2 +
 \frac{\tau^4}{15} \left[(\textbf{b}(\mbox{\boldmath$\nu$}
\mbox{\boldmath$\nabla$})^2 \textbf{b}) 
+\frac{1}{3} ((\mbox{\boldmath$\nu$}
\mbox{\boldmath$\nabla$})\textbf{b})^2 \right]  \right\rbrace, 
\label{3.2}
\end{equation} 
where $\textbf{b}=\mbox{\boldmath$\nabla$}U(\mbox{\boldmath$\varrho$})/m,~
\mbox{\boldmath$\nabla$}=\partial/\partial \mbox{\boldmath$\varrho$}$.
For the pre-exponential factor in Eq.(\ref{10.1}) we find 
\begin{equation}
[\ldots] \simeq \varphi_1(\xi)-\varphi_2(\xi)\tau^2\left[\textbf{b}^2 +
\frac{\tau^2}{3} (\textbf{b}(\mbox{\boldmath$\nu$}
\mbox{\boldmath$\nabla$}) \textbf{b}) \right] 
\label{4.2}
\end{equation}
Taking the integral over $\tau$ we obtain the spectral intensity
for $\vartheta_0 \ll V_0/m$.
\begin{eqnarray}
&& dI^F_{\xi}(\omega)=\frac{\alpha m^2\omega d\omega}{2\sqrt{3}\pi\varepsilon^2}
\int \frac{d^2\varrho}{s} \Bigg\{\int \frac{d^2\varrho_0}
{s(\varepsilon_{\perp}(\mbox{\boldmath$\varrho$}_0))}
\vartheta((\varepsilon_{\perp}(\mbox{\boldmath$\varrho$}_0)-
U(\mbox{\boldmath$\varrho$})) R_0(\lambda)
\nonumber \\
&& -\frac{(\textbf{b}(\mbox{\boldmath$\nu$}
\mbox{\boldmath$\nabla$})^2 \textbf{b})}{3\textbf{b}^4}
R_1(\lambda)+
\frac{\lambda}{30\textbf{b}^4}\left[((\mbox{\boldmath$\nu$}
\mbox{\boldmath$\nabla$})\textbf{b})^2 +3 (\textbf{b}(\mbox{\boldmath$\nu$}
\mbox{\boldmath$\nabla$})^2 \textbf{b})\right]R_2(\lambda)\Bigg\}, 
\label{5.2}
\end{eqnarray} 
where 
\begin{eqnarray}
\hspace{-15mm}&& R_0(\lambda)=\varphi_2(\xi)K_{2/3}(\lambda)-
\varphi_1(\xi)\int_{\lambda}^{\infty} K_{1/3}(y)dy, 
\nonumber \\
&&R_1(\lambda)=
\varphi_2(\xi)\left(K_{2/3}(\lambda)-\frac{2}{3\lambda}K_{1/3}(\lambda)\right),
\nonumber \\
\hspace{-15mm}&& R_2(\lambda)=\varphi_1(\xi)\left( K_{1/3}(\lambda) 
-\frac{4}{3\lambda}K_{2/3}(\lambda)\right)
\nonumber \\
&&+\varphi_2(\xi)\left(\frac{4}{\lambda}K_{2/3}(\lambda) 
-\left(1+\frac{16}{9\lambda^2} \right)K_{1/3}(\lambda) \right),
\label{6.2}
\end{eqnarray} 
here $\displaystyle{\lambda=\frac{2m^2\omega}{3\varepsilon\varepsilon'|\textbf{b}|}}$,
$K_{\nu}(\sigma)$ is the modified Bessel function (McDonald's function). 
Since the expression for $dI^F$ is independent of $z$, it follows that 
$\int d^3r/V \rightarrow \int d^2\varrho/s$, where $s$ is the transverse cross
section area per axis. The term in Eq.(\ref{5.2}) with $R_0(\lambda)$
represent the spectral intensity in the magnetic bremsstrahlung limit
with the flux redistribution taken into account. The other terms are the 
correction proportional $\vartheta_0^2$.

If the potential $U(\mbox{\boldmath$\varrho$})$ can be considered as
axially symmetric, we put  $U=U(\mbox{\boldmath$\varrho$}^2)$ and one 
can integrate over angles of vector $\mbox{\boldmath$\varrho$}$. We
obtain
\begin{eqnarray}
&& dI_{\xi}^F(\omega)=\frac{\alpha m^2\omega d\omega}{2\sqrt{3}\pi\varepsilon^2}
\int_{0}^{x_0} \frac{dx}{x_0} \Bigg\{\int_{0}^{x_0} \frac{dy}
{y_0(\varepsilon_{\perp}(y))}
\vartheta((\varepsilon_{\perp}(y)-
U(x)) R_0(\lambda)
\nonumber \\
&& -\frac{1}{6}\left( \frac{m \vartheta_0}{V_0}\right)^2\Bigg[
\frac{xg''+2g'}{xg^3} R_1(\lambda)
\nonumber \\
&& -\frac{\lambda}{20g^4x^2}\left(2x^2g^{\prime 2}+g^2+14gg^{\prime}x
+6x^2gg^{\prime \prime}\right)R_2(\lambda)\Bigg] \Bigg\}, 
\label{7.2}
\end{eqnarray} 
where we have adopted a new variable $x=\mbox{\boldmath$\varrho$}^2/a_s^2,
~x \leq x_0,~x_0^{-1}=\pi a_s^2dn_a=\pi a_s^2/s,~a_s$ 
is the effective screening radius
of the potential of the string, $n_a$ is the density of atoms in a crystal,
$d$ is the average distance between atoms of a chain forming the axis;
\begin{equation}
\varepsilon_{\perp}(y)=\frac{\varepsilon\vartheta_0^2}{2}+U(y),\quad
y_0(\varepsilon_{\perp}(y))=\int_{0}^{\infty}
\vartheta(\varepsilon_{\perp}(y)-U(x))dx.
\label{8.2}
\end{equation} 
The notation $U^{\prime}(x)=V_0g(x)$ is used in Eq.(\ref{7.2}) and
\begin{equation}
\lambda=\frac{m^3a_s\omega}{3\varepsilon\varepsilon'V_0g(x)\sqrt{x}}=
\frac{u}{3\chi_s g(x)\sqrt{x}},~\chi_s=\frac{V_0\varepsilon}{m^3 a_s},
~u=\frac{\omega}{\varepsilon'}
\label{9.2}
\end{equation} 
For specific calculation we use the following for the potential of axis:
\begin{equation}
U(x)=V_0\left[\ln\left(1+\frac{1}{x+\eta} \right)- 
\ln\left(1+\frac{1}{x_0+\eta} \right) \right]. 
\label{10.2}
\end{equation} 
For estimates one can put $V_0 \simeq Ze^2/d,~\eta \simeq 2u_1^2/a_s^2$,
where $Z$ is the charge of the nucleus, 
$u_1$ is the amplitude of thermal vibrations,
but actually the parameter of potential were determined by means of a fitting
procedure using the potential Eq.(\ref{8.1}) (table of parameters for 
different crystals is given in Sec.9 of \cite{BKS}).
For this potential
\begin{equation}
g(x)=\frac{1}{x+\eta} - \frac{1}{x+\eta +1} = \frac{1}{(x+\eta)(x+\eta +1)}
\label{11.2}
\end{equation} 

We shall assume, for the sake of simplicity, that the distribution is uniform 
over the transverse coordinates. This is true in the case of large angles of
incidence $\vartheta_0 > \vartheta_c$ and is approximately correct in the case
of beams with a large angular spread $\Delta \vartheta_0 \sim \vartheta_c$.
In this case and for the assumptions 
$u \ll \chi_s,~(\chi_s/u)^{4/3}/\varrho(\vartheta_0)\ll 1$ 
we have 
\begin{equation}
\frac{dI_{\xi}^M}{\omega d\omega}=\frac{\alpha m^2}{6\sqrt{3}\pi x_0 \varepsilon^2}
\Phi_{\xi}
\label{12.2}
\end{equation} 
where 
\begin{equation}
\Phi_{\xi}=\Gamma\left(\frac{2}{3}\right)\left(\frac{6\chi_s}{u}\right)^{2/3}
\left[\varphi_2(u)\left(\ln\frac{18\sqrt{3}\chi_s}{u} - \frac{\pi}{2\sqrt{3}}-
C -l_1(\eta)\right)-\frac{3}{2}\varphi_1(u)\right]
 \label{13.2}
\end{equation} 
here
\begin{eqnarray}
&&\varphi_1(u)=1+\frac{\xi u}{1+u}, \quad \varphi_2(u)=(1+\xi)(1+u)+\frac{1-\xi}{1+u},
\quad C=0.577216..., 
\nonumber \\
&& l_1(\eta)=3.975 \beta^{2/3}\left(1+\frac{8}{15}\beta 
+\frac{7}{18}\beta^2\right)-\beta\left(\frac{3}{2}+\frac{9}{8}\beta 
+\frac{13}{14}\beta^2\right),
\label{14.2}
\end{eqnarray} 
where $\beta=\eta/(1+\eta)$. In derivation of Eq.(\ref{12.2}) the integration
over $x$ is carried out in the interval $(0, \infty)$. Since the main (logarithmic)
contribution into the integral in Eq.(\ref{7.2}) comes from the interval $x \sim 
(\chi_s/u)^{2/3}$, it is clear that the asymptotic expression Eq.(\ref{12.2})
 is valid only is $(\chi_s/u)^{2/3} < x_0$.

At high energies ($\chi_s \gg 1$) the expression for the intensity
of radiation in the magnetic bremsstrahlung limit 
$I_{\xi}^M(\varepsilon)$ becomes
\begin{equation}
I_{\xi}^M(\varepsilon) \simeq \frac{\alpha V_0 g_1\varepsilon}{2m x_0 a_s \chi_s^{1/3}}
\left[\left(1+\frac{11}{16}\xi\right)(\ln  \chi_s +g_0 )- \frac{231}{512}\xi\right],
\label{15.2}
\end{equation} 
where
\begin{eqnarray}
&& g_1=\left(\frac{2}{3}\right)^6 6^{2/3}\Gamma\left(\frac{2}{3}\right)\simeq 0.3925
\nonumber \\
&& g_0=\frac{\pi}{2\sqrt{3}}+\frac{5}{2}\ln 3 +\ln 2 -C -l_1(\eta) 
\simeq 0.6756-l_1(\eta).  
\label{16.2}
\end{eqnarray}

The circular polarization of radiation is defined by Eq.(\ref{7.1a}),
where we substitute $dE_{\pm} \rightarrow dI_{\pm}$. In the hard end of spectrum 
$(\varepsilon' \ll \varepsilon)~\Lambda \simeq 1$ 
since $dI_+ \propto \varepsilon/\varepsilon'$ and 
$dI_- \propto \varepsilon'/\varepsilon$. 
In the limit $\chi_s \gg 1$ the radiation spectrum extends into hard part 
up to $u=\omega/(\varepsilon- \omega) \sim \chi_s a_s/u_1$, where under condition
$|(\mbox{\boldmath$\zeta$}{\bf v})| \simeq 1$ the emitted photons have
practically the complete circular polarization (the  helicity coincide with
electron helicity).    
For the integral radiation intensity (where the main contribution gives
the region $\omega \sim \varepsilon-\omega$) the circular polarization
follows from Eq.(\ref{15.2}):
\begin{equation}
\Lambda = \frac{11}{16}\left(1- \frac{21}{32(\ln \chi_s + g_0)}\right).   
\label{18.2}
\end{equation} 
Here the main term coincide with circular polarization of radiation
of polarized electrons in external electromagnetic field, see Eq.(4.88)
in \cite{BKS}.

In Fig.1 the spectral intensities of radiation 
$dI^M_+/d\omega$,$~dI^M_-/d\omega$ and the 
spectral intensity of radiation of unpolarized electrons 
$dI^M/d\omega=dI^M_+/d\omega + dI^M_-/d\omega$ in tungsten crystal,
axis $<111>$, 
temperature $T=293~K$ are plotted for energy $\varepsilon=$250~GeV
(curves 1,2,3, respectively) and for energy $\varepsilon=$1~TeV 
(curves 4,5,6, respectively). It is seen that when $\omega/\varepsilon 
\rightarrow 1$ the intensity $dI^M_+/d\omega$ dominates 
($dI^M_+/d\omega \gg dI^M_-/d\omega$), while for 
$\omega/\varepsilon \ll 1$ the degree of polarization diminishes 
($dI^M_+/d\omega \rightarrow dI^M_-/d\omega$). 
It should be noted that Eq.(\ref{7.2a}) is 
fulfilled within  accuracy better 10\% for $\omega/\varepsilon \leq 0.2$.
In Fig.2 the circular polarization $\xi^{(2)}$ of radiation is plotted 
versus $\omega/\varepsilon$ for the same crystal. This curve is true
for both energies:$\varepsilon=$250~GeV and  $\varepsilon=$1~TeV. 
Actually this means that it is valid for any energy in high-energy region.
At $\omega/\varepsilon$=0.8 one has $\xi^{(2)}$=0.94 and at 
$\omega/\varepsilon$=0.9 one has $\xi^{(2)}$=0.99.

It is instructive to compare the obtained results with bremsstrahlung of
polarized electrons in amorphous medium which one write as (see e.g.\cite{BKF})
\begin{equation}
\varepsilon \frac{dI_{\xi}^{C}}{d\omega}= 
\frac{\alpha m^2 \varepsilon}{8 \pi \varepsilon_e}\left[
(1+\xi)\left(1-\frac{2}{3}\frac{\varepsilon'}{\varepsilon}\right)
+\frac{\varepsilon'^2}{\varepsilon^2}\left(1-\frac{\xi}{3}\right)
+\frac{2}{9L_1}\frac{\varepsilon'}{\varepsilon}
\left(1+\xi\frac{\omega}{\varepsilon}\right)\right], 
\label{19.2}
\end{equation} 
where $\varepsilon_e=m(8\pi Z^2\alpha^2 n_a \lambda_c^3L_1)^{-1}$, 
for notations see Eqs.(\ref{7.2}) and (\ref{10.2}), $\lambda_c=1/m$
is the electron Compton wavelength, $L_1=2(\ln 183Z^{1/3} - f(Z\alpha)),~
f(z)={\rm Re}[\psi(1+iz)-\psi(1)]$, $\psi(x)$ is the logarithmic 
derivative of the gamma function, the function $f(Z\alpha)$ gives
the Coulomb corrections, for tungsten
$L_1$=6.99, $\varepsilon_e=2.73~$TeV. The dependence of the circular
polarization $\xi^{(2)}$ on $\omega/\varepsilon$ for the
intensity Eq.(\ref{19.2}) is very close to shown in Fig.2. 
For the integral bremsstrahlung intensity Eq.(\ref{19.2}) 
one has $\Lambda=5/9$.
However, as one can be seen from Fig.1, in the region 
$\omega \sim \varepsilon$ the magnitude of intensity
in oriented crystal is around 10 times larger than given by
Eq.(\ref{19.2}) for both energies:$\varepsilon=$250~GeV and  
$\varepsilon=$1~TeV.
The value $r=(I^M/I^C)_{max}= (L_{rad}/L_{ch})_{max}$ for different
crystals is discussed in Sec.17.3 and is given in Table 17.1 in
\cite{BKS}. This value attains maximum value for light elements: 
in diamond for axis $<111>$ $r$=168, in silicon for axis $<110>$ $r$=81.

It should be mentioned that for energies $\varepsilon \geq \varepsilon_e$
the process of hard photon emission ($\omega \sim \varepsilon$) is 
affected by the multiple scattering (the LPM effect). At 
$\varepsilon \gg \varepsilon_e$ the suppression of spectral intensity
of bremsstrahlung become essential for the whole spectrum 
\cite{BK3}, \cite{BK}. 
On the contrary, the influence of multiple scattering on the photon
emission in a oriented crystal is rather weak because the formation length
in a oriented crystal is much shorter than in an amorphous medium 
(see Sec. 21 in \cite{BKS}).

\section{Modified theory of coherent bremsstrahlung}
\setcounter{equation}{0}

The estimates of double sum in the phase $A_2$ made at the beginning of previous 
section: $\sim (\vartheta_v/\vartheta_0)^2\Psi$ remain valid also for 
$\vartheta_0 \geq \vartheta_v$, except that now the factor in the double sum is
$(\vartheta_v/\vartheta_0)^2 \leq 1$, so that the values 
$|q_{\parallel}\tau| \sim 1$ contribute. We consider first the limiting case
$\vartheta_0 \gg \vartheta_v$, then this factor is small and $\exp(-iA_2)$
can be expanded accordingly. As a result Eq.(\ref{10.1}) acquires the
form
\begin{eqnarray}
&& dI_{\xi}^{coh}(\omega)=-\frac{i\alpha m^2\omega d\omega}{4\pi\varepsilon^2}
\int_{-\infty}^{\infty} \frac{d\tau}{\tau-i0} 
\exp\left(-i\frac{m^2\omega\tau}{\varepsilon\varepsilon'} \right)
\sum_{\textbf{q},\textbf{q}^{\prime} }\frac{G(\textbf{q})G(\textbf{q}')}
{m^2 q_{\parallel} q_{\parallel}'}
(\textbf{q}_{\perp}\textbf{q}_{\perp}')
\nonumber \\
&& \left[\varphi_2(\xi)\sin(q_{\parallel}\tau) \sin(q_{\parallel}'\tau)
+i\varphi_1(\xi)\frac{m^2\omega\tau}{\varepsilon\varepsilon'} 
\Psi(q_{\parallel}, q_{\parallel}', \tau)\right] \int \frac{d^3r}{V} \exp [-i(\textbf{q}+\textbf{q}')\textbf{r}].
\label{1.3}
\end{eqnarray} 
The integration over coordinate $\textbf{r}$ in Eq.(\ref{1.3}) is elementary and
gives $\delta_{\textbf{q}+\textbf{q}', 0}$, after which the sum over $\textbf{q}'$
and the integrals over $\tau$ are easily calculated by means of the theory of
residues. Finally we obtain
\begin{eqnarray}
&& dI_{\xi}^{coh}(\omega)=\frac{\alpha \omega d\omega}{8\varepsilon^2}
\sum_{\textbf{q}} |G(\textbf{q})|^2 \frac{\textbf{q}_{\perp}^2}{q_{\parallel}^2}
\left[\varphi_2(\xi) -\varphi_1(\xi)\frac{2m^2\omega}
{\varepsilon\varepsilon'q_{\parallel}^2}\left(|q_{\parallel}|- 
\frac{m^2\omega}{2\varepsilon\varepsilon'}\right)  \right] 
\nonumber \\
&& \times \vartheta\left(|q_{\parallel}|- 
\frac{m^2\omega}{2\varepsilon\varepsilon'}\right). 
\label{2.3}
\end{eqnarray} 
The spectral distribution Eq.(\ref{2.3}) coincide with the result of standard
theory of coherent bremsstrahlung (CBS), see e.g. \cite{TM}. 

In the case $\chi_s \gg 1$ ($\chi_s$ is defined in Eq.(\ref{9.2})), one can
obtain from general expression Eq.(\ref{10.1}) the expression for spectral
intensity, the region of applicability of which is broader than that of
standard CBS theory. For this purpose it is necessary to take into account that
the phase $A_2$ Eq.(\ref{11.1}) has for $q_{\parallel}+ q_{\parallel}' \neq 0$
terms of the order $(\vartheta_v/\vartheta_0)^3/\chi_s$ and 
$(\vartheta_v/\vartheta_0)^4/\chi_s^2$ which can be small even for 
$\vartheta_0 \leq\vartheta_v$ if $\chi_s \gg 1$. Therefore, assuming that these
contributions are small, we carry out the corresponding expansion of $\exp(-iA_2)$,
while the term with $q_{\parallel}+ q_{\parallel}' = 0$ in the double sum in 
$A_2$ will be retained in the exponent. As a result we obtain an expression 
which coincides in the form with Eq.(\ref{1.3}) where we must make the 
substitution
\begin{equation}
\exp\left(-i\frac{m^2\omega\tau}{\varepsilon\varepsilon'} \right) \rightarrow
\exp\left(-i\frac{m_{\ast}^2\omega\tau}{\varepsilon\varepsilon'} \right), \quad
m_{\ast}^2=m^2\left(1+\frac{\varrho}{2} \right). 
\label{3.3}
\end{equation} 
Above the parameter $\varrho$ (Eq.(\ref{1.1})) has the form
\begin{equation}
\frac{\varrho}{2}=\frac{1}{m^2}\sum_{\textbf{q},\textbf{q}'}G(\textbf{q})
G(\textbf{q}')
\frac{\textbf{q}_{\perp}\textbf{q}_{\perp}'}{q_{\parallel} q_{\parallel}'}
\left[\delta_{q_{\parallel}+ q_{\parallel}',0} -\delta_{q_{\parallel},0}
\delta_{q_{\parallel}',0}\right] =\sum_{\textbf{q},q_{\parallel} \neq 0}
\frac{|G(\textbf{q})|^2\textbf{q}_{\perp}^2 }{m^2q_{\parallel}^2},
\label{4.3}
\end{equation} 
and in the term $\displaystyle{\frac{\sin(q_{\parallel}+ q_{\parallel}')\tau}
{(q_{\parallel}+ q_{\parallel}')\tau}}$ it is necessary to assume that
$q_{\parallel}+ q_{\parallel}' \neq 0$. The remaining calculations 
are carried out in the same way as in the transition from Eq.(\ref{1.3}) to
Eq.(\ref{2.3}). The final result can be presented in a form 
\begin{eqnarray}
&& \frac{dI_{\xi}^{mcoh}(\omega)}{d\omega}=\frac{\alpha u}{8\varepsilon}
\sum_{\textbf{q}} |G(\textbf{q})|^2 \frac{\textbf{q}_{\perp}^2}{q_{\parallel}^2}
\Bigg[1+\xi +\frac{1-\xi}{(1+u)^2} 
\nonumber \\
&& -\frac{8[1+u(1+\xi)]}{(2+\varrho)(1+u)^2}\frac{u}{u_0}\left(1- 
\frac{u}{u_0}\right) \Bigg] \vartheta(u_0-u), 
\label{5.3}
\end{eqnarray}
where
\begin{equation}
u_0=\frac{4\varepsilon q_{\parallel}}{m^2 (2+\varrho)}, \quad
u=\frac{\omega}{\varepsilon'}
\label{5.3a}
\end{equation}  
Equation (\ref{5.3}) is not more complicated than Eq.(\ref{2.3})
but has a significantly broader range of applicability.

The spectral intensities Eqs.(\ref{2.3}) and (\ref{5.3}) can be
much higher than the Bethe-Heitler bremsstrahlung intensity for 
small angles of incidence $\vartheta_0$ with respect to selected axis.
For the case $\vartheta_0 \ll 1$ the quantity $q_{\parallel}$
can be represented as
\begin{equation}
q_{\parallel} \simeq \frac{2\pi}{d}n+\textbf{q}_{\perp}\textbf{v}_{0.\perp}  
\label{5.3b}
\end{equation} 
The main contribution to the sum in Eqs.(\ref{2.3}) and (\ref{5.3}) for
small $\vartheta_0$ is given $\textbf{q}$ with $n=0$, then 
\begin{equation}
q_{\parallel} \simeq \left(\frac{2\pi}{f}k \cos \varphi +  
\frac{2\pi}{h}l \sin \varphi\right) \vartheta_0,
\label{5.3c}
\end{equation} 
where $f$ and $h$ are the characteristic periods of the potential in the
transverse plane.

Let us consider the spectral intensity of radiation in the extreme limit,
when the parameter $\lambda \equiv 2\varepsilon|q_{\parallel}|_{min}/m^2
\sim \varepsilon \vartheta_0/m^2 a_s \gg 1$. In this case the maximum 
of intensity of coherent bremsstrahlung is attained at such values of
$\vartheta_0$ where the standard theory of coherent bremsstrahlung
becomes invalid. Bearing in mind that if $\lambda \gg 1$  and 
$\vartheta_0 \sim V_0/m$, then $\chi_s \sim \lambda \gg 1$,
we can conveniently use a modified theory of coherent bremsstrahlung.

The direction of transverse components of particle's velocity in 
Eq.(\ref{5.3b}) can be selected in a such way, that
the spectral intensity described by Eq.(\ref{5.3}) has a sharp maximum
near the end of spectrum at 
$\omega = 2\varepsilon\lambda(2+2\lambda+\varrho)^{-1}$
with relatively small (in terms of $\lambda^{-1}$) width 
$\Delta \omega \sim \varepsilon(1+\varrho/2)/\lambda=
m^2(1+\varrho/2)/2 |q_{\parallel}|_{min}$:
\begin{equation}
\left(\frac{dI_{\xi}}{d\omega}\right)_{max}=
\frac{\alpha \varrho |q_{\parallel}|_{min}}{4(2+\varrho)}
\left(1+\xi + \frac{1-\xi}{(1+u_m)^2}\right),\quad 
u_m=\frac{2\lambda}{2+\varrho}.
\label{8.3}
\end{equation} 
It is seen that in the maximum of spectral distribution the radiation 
intensity with opposite helicity ($\xi=-1$) is suppressed as 
$1/(1+u_m)^2$. At $u > u_m$ one have to take into account the next
harmonics of particle acceleration. In this region of spectrum 
the suppression of  radiation intensity with opposite helicity is more strong,
so the emitted photons have nearly complete circular polarization.

Comparing Eq.(\ref{8.3}) with Eq.(\ref{19.2}) for bremsstrahlung
for $\varepsilon' \ll \varepsilon$ we find that for the same circular 
polarization ($\xi^{(2)} \simeq (\mbox{\boldmath$\zeta$}{\bf v})$) in the 
particular case $\varrho = 1$ the magnitude of spectral intensity in 
Eq.(\ref{8.3}) is about $\chi_s(\varepsilon_e)$ times larger than
the spectral intensity in Eq.(\ref{19.2}). For tungsten W one has
$\chi_s(\varepsilon_e)=78$. Taking into account that $\Delta \omega
\sim \varepsilon/\chi_s(\varepsilon) =
\varepsilon_e/\chi_s(\varepsilon_e)$ we find that
\begin{equation}
\Delta \omega  \left(\frac{dI_{\xi}}{d\omega}\right)_{max}
\sim \frac{\varepsilon_e}{L_{rad}},\quad 
\frac{dN_{\gamma}}{dt}
\sim \frac{\varepsilon_e}{\varepsilon}\frac{1}{L_{rad}},
\label{9.3}
\end{equation} 
where $N_{\gamma}$ is the number of emitted photons.
So, it is seen from above analysis that under these conditions the
considered mechanism of emission of photons with circular polarization
is especially effective because there is gain both in monochromacity
of radiation and total yield of polarized photons near 
hard end of spectrum.

\section{Conclusions}
\setcounter{equation}{0}

It is shown above that at high energy the radiation from 
longitudinally polarized electrons in oriented crystals is circularly
polarized and $\xi^{(2)} \rightarrow 1$ near the end of spectrum.
This is true in magnetic bremsstrahlung limit $\vartheta_0 \ll V_0/m$
as well as in coherent bremsstrahlung region $\vartheta_0 > V_0/m$. 
This is particular case of helicity transfer.

It should be noted that in crossing channel: production of 
electron-positron pair with longitudinally polarized particles
by the circularly polarized photon in an oriented crystal the same phenomenon
of helicity transfer takes place in the case when the final particle 
takes away nearly all energy of the photon. The corresponding formulas can be
obtained from given above using substitution rules. 
For processes in external electromagnetic 
field this item is discussed in Sec.3.3 of \cite{BKS}. 

So, the oriented crystal is a very effective device for helicity transfer 
from an electron to photon and back from a photon to electron or positron.
Near the end of spectrum this is nearly 100\% effect.

\vspace{0.5cm}

{\bf Acknowledgments}

The authors are indebted to the Russian Foundation for Basic
Research supported in part this research by Grant 
03-02-16154.

\newpage

{\bf Figure captions}

\vspace{8mm}
\begin{itemize}

\item {\bf Fig.1} Spectral intensity of radiation 
$\displaystyle{\varepsilon\frac{dI^M(\omega)}{d\omega}}$ in tungsten crystal, 
axis $<111>$, $T=293~K$ in units $\alpha m^2$ vs $\omega/\varepsilon$. For the
initial electron energy $\varepsilon$=250~GeV curve 1 is $dI^M_-/d\omega$,
curve 2 is $dI^M_+/d\omega$, curve 3 is $dI^M_+/d\omega+dI^M_-/d\omega$;
and for the
initial electron energy $\varepsilon$=1~TeV curve 4 is $dI^M_-/d\omega$,
curve 5 is $dI^M_+/d\omega$, curve 6 is $dI^M_+/d\omega+dI^M_-/d\omega$.

\item {\bf Fig.2} The circular polarization $\xi^{(2)}$ 
of radiation (for ($\mbox{\boldmath$\zeta$}{\bf v}$)=1)
vs $\omega/\varepsilon$ for the tungsten crystal, 
axis $<111>$, $T=293~K$. The curve is valid for both energies:
$\varepsilon$=250~GeV and $\varepsilon$=1~TeV.

\end{itemize}

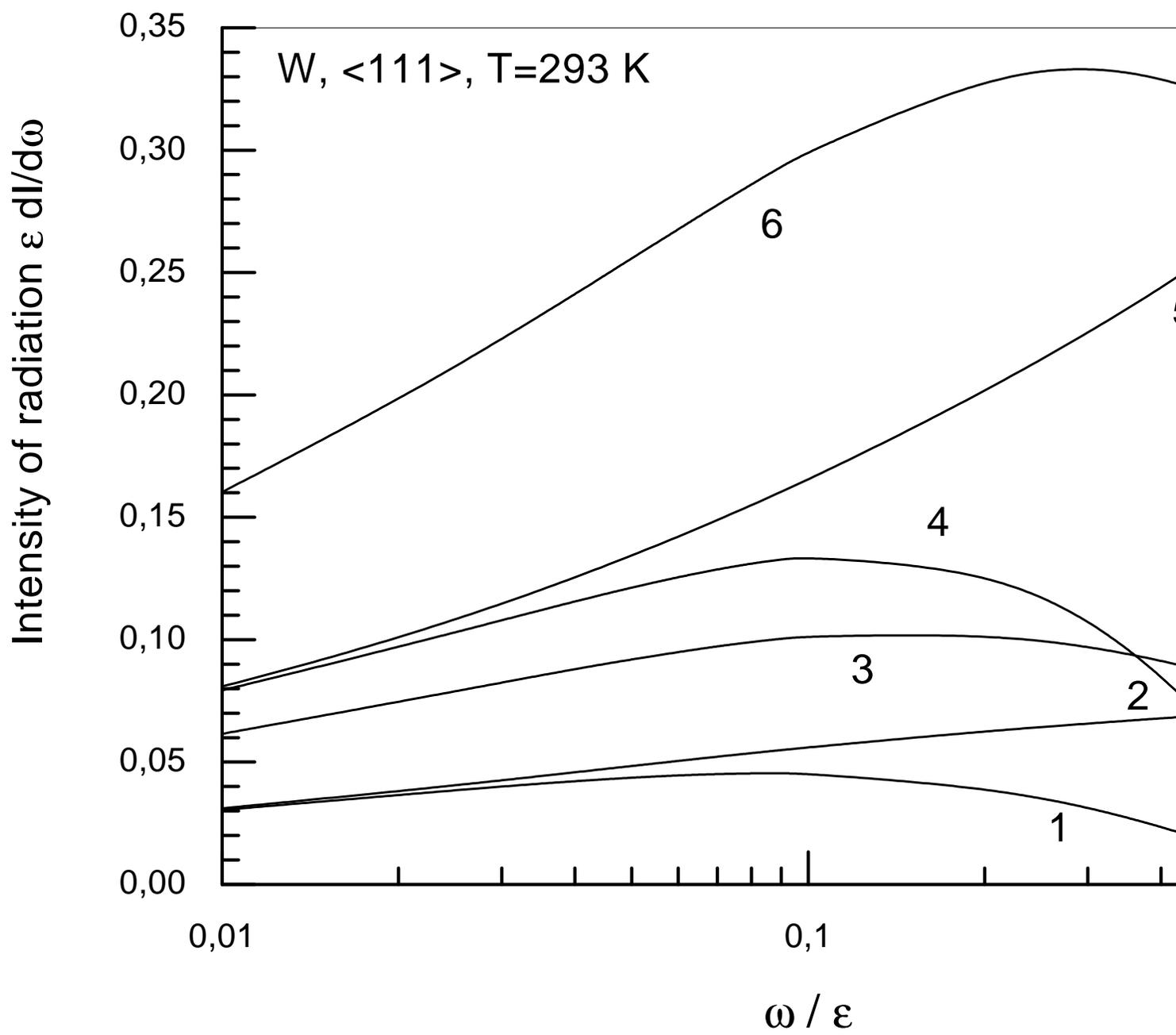
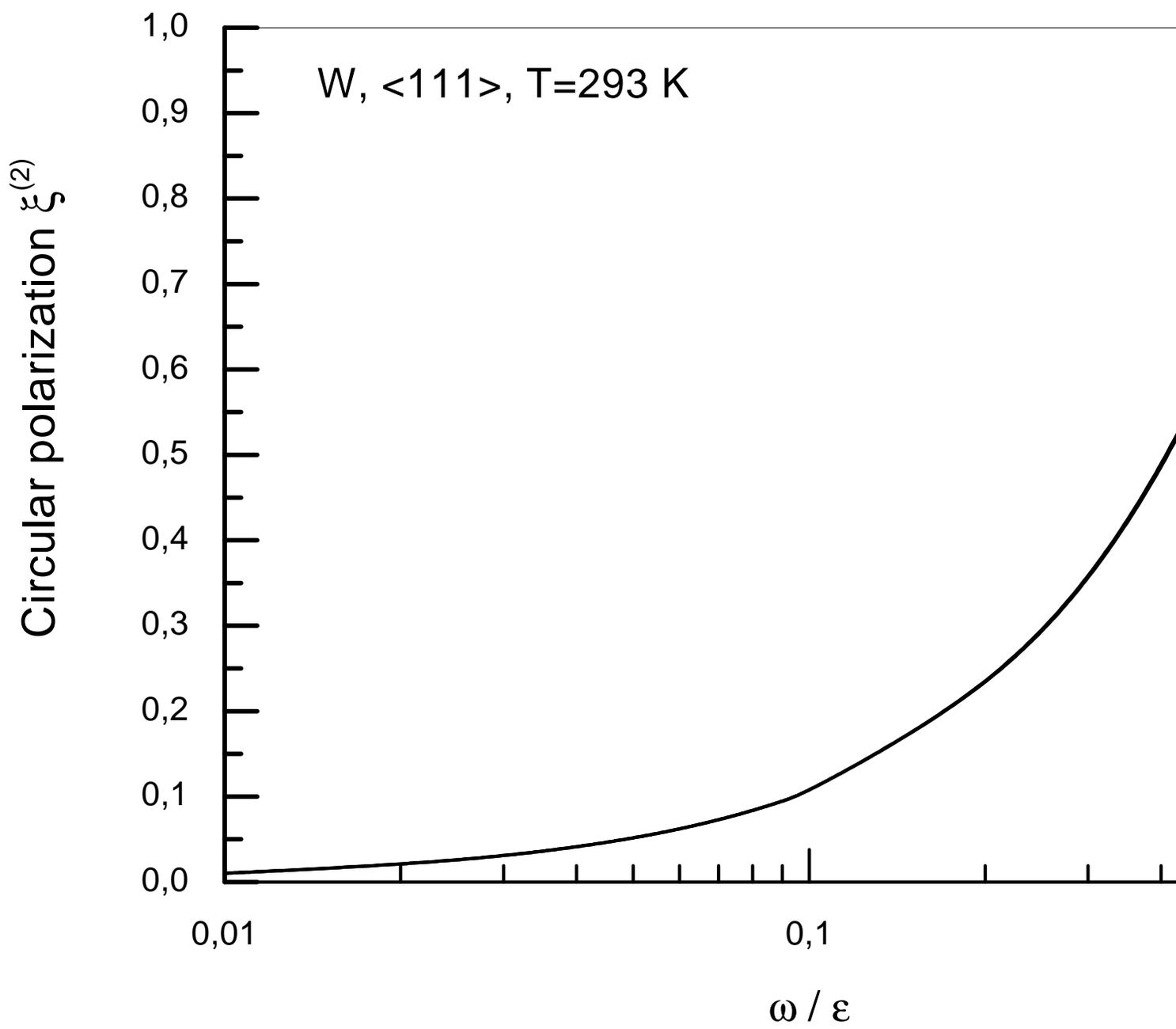
\end{document}